\documentclass[aps,prl,reprint,superscriptaddress]{revtex4-2}

\usepackage{graphicx}
\usepackage{amssymb,amsmath,amsfonts,latexsym,mathrsfs,amsthm}
\usepackage{stmaryrd}
\usepackage{soul}
\usepackage{color}

\newcommand{\CuAlMn}{Cu$_{67}$Al$_{24}$Mn$_{9}$ }

\sethlcolor{yellow}

\begin{document}

\title{Bimodal Scaling Law and Size Effect In Superelastic Nanopillars}

\author{Mostafa Karami}
\affiliation{Department of Mechanical and Aerospace Engineering, Hong Kong University of Science and Technology, Clear Water Bay, Hong Kong}
\affiliation{Department of Materials Science and Engineering, Sharif University of Technology, Tehran, Iran}
\author{Xian Chen}
\email{xianchen@ust.hk}
\affiliation{Department of Mechanical and Aerospace Engineering, Hong Kong University of Science and Technology, Clear Water Bay, Hong Kong}

\begin{abstract}

Shape memory alloys that can deform and then spring back to their original shape, have found a wide range of applications in the medical field, from heart valves to stents. As we push the boundaries of technology creating smaller, more precise tools for delicate surgery treatments, the behavior of these alloys at tiny scales becomes increasingly crucial. In this study, we discover that the size effect of critical stress required for stress-induced phase transformation is not universal. We propose an orientation-dependent power decay law, indicating a specific increase in critical stress for pillars smaller than 1 micron meter for the nominally soft [001] and hard [111] orientations. Additionally, we observe high transformability with 11\% recoverable strain under high stress (2 GPa) through lattice frustration at 200nm scale. This research opens new avenues for exploring the superior elastic behavior of shape memory alloys for nanodevices.

\end{abstract}

\maketitle

Shape memory alloys, characterized by their large deformability under moderate stress and the ability to revert to their original shapes upon stress removal, have demonstrated their success in many biomedical applications such as artificial heart valves, vascular stents and other biocompatible implants \cite{levi_2008,jansen_1992, chen_2007rapidly,yeazel_2020, bahraminasab_2013}. As the demand escalates for miniaturized implants tailored for the treatment of neurological diseases and microsurgical tools\cite{bechtold_2016method, bechtold2016fab}, the superelastic properties of shape memory alloys exhibited at micro-to-nanoscales become increasingly pertinent. The size-dependent functionality and reversibility of these materials are garnering considerable attention in the realm of nanotechnology.

Nanomechanics experiments have empirically demonstrated that the critical stress required for stress-induced transformation in shape memory alloys increases as the sample size diminishes. \cite{frick_2007, ozdemir2012size, gomez_2017,frick_2010} Concurrently, however, the prominence of superelastic strain gradually recedes as dimensions reduce from micron to nano meters.  For instance, in NiTi compression pillars, the superelasticity becomes less distinct below 400 nm, and disappears entirely around 200 nm.\cite{frick_2007,frick_2010} A similar trend was observed in single crystal Ni$_{54}$Fe$_{19}$Ga$_{27}$ nanopillars with diameters less than 400 nm.\cite{ozdemir2012size} Furthermore,  many nanocrystalline NiTi-based alloys exhibit a weakened superelastic feature accompanied by a significantly narrowed hysteresis loop during stress-induced phase transformation.\cite{ahadi2014effects, xiao2017grain, hua_2018} The size effect in these alloys was typically observed in grains measuring around 20 nm, which is considerably smaller than the size effect seen in single-crystal nanocompression experiments. In some case, small grains under plastic deformation may induce nano-sized amorphous clusters, still showing magnificent recoverable strain upon 2 GPa compressive stress.\cite{hua_2021nat}, which makes the NiTi alloy attractive under extreme loading conditions.  Unlike NiTi alloys, the single crystal [001] oriented CuAlNi nanopillars exhibit definitive superelasticity even around 260 nm corresponding to a wide open stress hysteresis under moderate stress-induced phase transformations. \cite{juan_2009nano, san2013superelasticity, gomez_2017} Superelastic transformability at a high stress has never been reported in CuAl-based shape memory micro/nanopillars. Moreover, the amorphization process and orientation dependent size effect are rarely reported for CuAl-based alloys. Grasping these effects is pivotal in revolutionizing the development of Cu-based polycrystals for small-scale applications.    

For most shape memory alloys, the scaling law associated with the critical stress required for phase transformation has been heuristically related to the plasticity model\cite{fuster_2020, dou2009universal} used in non-transforming metals \cite{jennings_2011, greer_2005,volkert_2006,dimiduk_2005, kim_2010, greer_2011rev, kim_2012}. It has been postulated that the critical stress for phase transformation bears a resemblance to the yield strength of ductile metals through a universal power law that scales as $\sigma_c \propto d^n$ with the exponent $n < 0$ for a wide dimension variation from 10 $\mu$m to 200 nm \cite{frick_2010,gomez_2017}. Empirically, the exponent of [210] oriented single crystal NiTi nanopillars was reported to be -0.23 \cite{frick_2010}, however no significant size correlation was observed for [001] and [111] oriented nanopillars. For CuAlNi, the exponent was fitted to be -2 for [001] oriented nanopillars  \cite{gomez_2017}, showing magnificently higher growth of critical stress with size decreasing. It is important to verify whether the scaling law of the critical stress with respect to the micro/nano pillar's size is generally applicable across different crystallographic orientations. Whether the universal power law is valid for the critical stress of shape memory alloys remains an open question.

To address these questions, in this paper, we will study the \CuAlMn nanopillars from 5 $\mu$m to 185 nm for different orientations by in situ nanomechanical compression experiments. The target alloy is counted as a relatively ductile shape memory alloy in the Cu-based alloy family. During both thermal-driven and stress-induced phase transformations, it forms typical laminated twin martensite from austenite through an elastic transition layer. From microstructural compatibility's point of view \cite{ball_1987,bhattacharya_2003}, this alloy is very similar to equal-atomic NiTi. With the microstructural affinity to the nitinol alloys, this alloy is a good candidate to be explored for high stress-induced transformability.   

\begin{figure}
    \centering
    \includegraphics[width=0.38\textwidth]{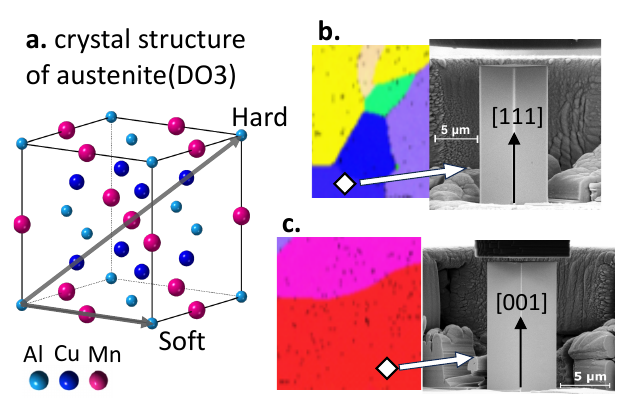}
    \caption{{\bf a.} DO3 structure ($Fm\bar 3 m$) of CuAlMn austenite. {\bf b.} The orientation map and the selected grain close to [111] orientation. {\bf c.} The orientation map and the selected grain close to [001] orientation. The cuboidal nanopillars were fabricated from these grains. We use $[001]$ and $[111]$ as the nominal orientations for these micro/nanopillars.\label{fig:modulus}}
\end{figure}

The \CuAlMn alloy shows substantial orientation dependency of superelasticity under both tensile and compressive stresses \cite{karami_2020two, karami_2022}. Besides, it exhibits a strong elastic anisotropy of austenite lattice. It allows for the possibilities of various deformation mechanisms co-activated during the stress-induced phase transformations. Methods for modulus determination is included in the Supplemental Materials \cite{SMref}. The maximum elastic modulus appears along [111] orientation, while the softest modulus is given by [001] orientation. According to the orientation map of the modulus in Fig. S1 \cite{SMref}, we chose the grains close to the softest ($\sim [001]$ pillar) and hardest ($\sim [111]$ pillar) orientations as illustrated in Fig. \ref{fig:modulus}. The uniaxial microcompression tests were conducted on the nominal [001] pillar with modulus 9 GPa and the nominal [111] pillar with modulus 65 GPa. Unlike traditional nanopillar used in the nanocompression experiments, we fabricated a series of zero-tapered cuboidal nanopillars with sizes varying from 5 $\mu$m to below 200 nm (Fig. S2 \cite{SMref}). The fabrication process is detailed in Methods \cite{SMref}. In such a way, the true stress-strain curves can be obtained without the influence of misalignment between the loading cell and tested pillars, and the true stress-stain behaviors can be accurately captured, especially for the submicron sized pillars. 

\begin{figure}
\centering
    \includegraphics[width=0.48\textwidth]{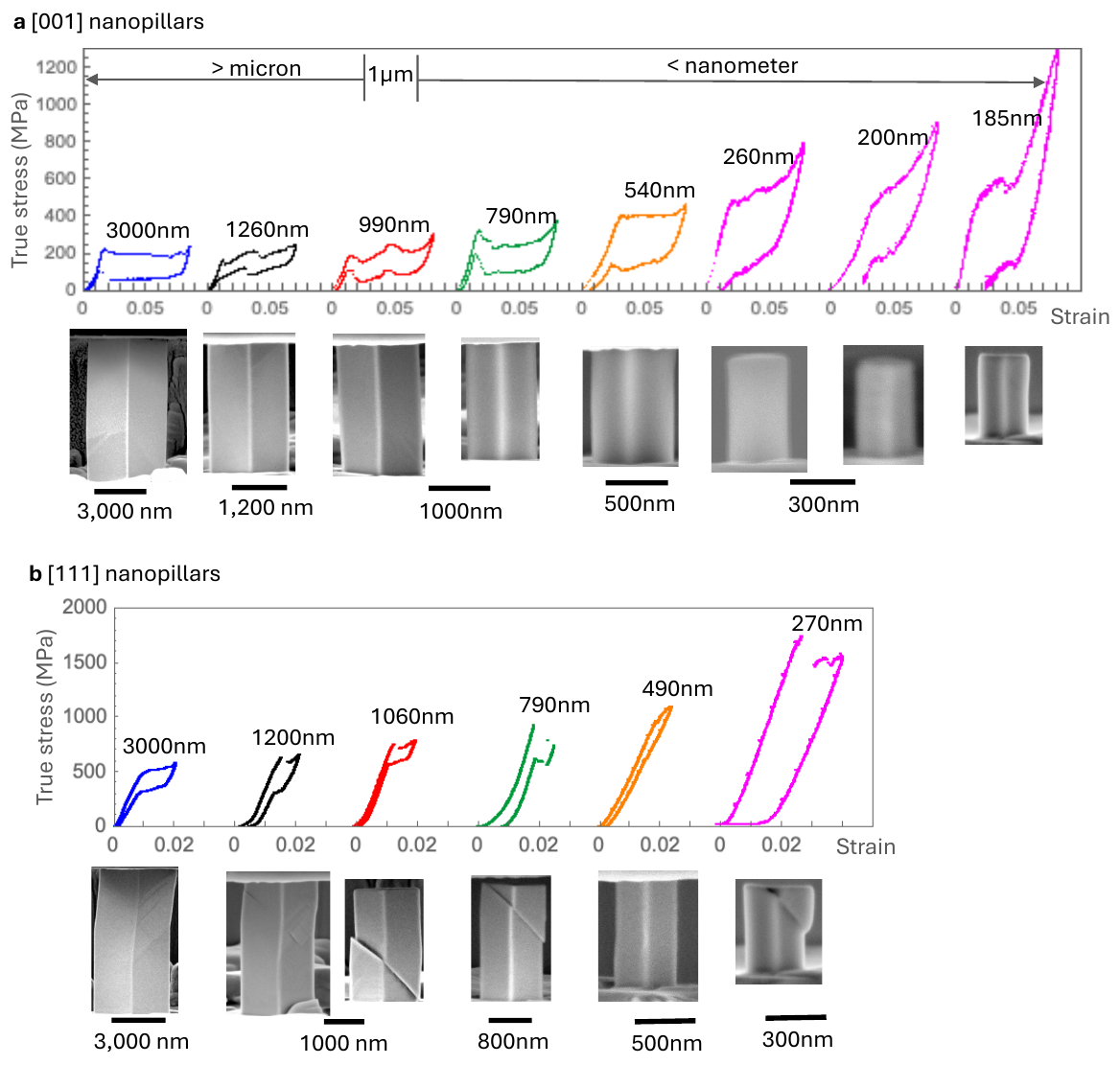}
    \caption{Stress-strain behaviors captured by uniaxial nanocompression tests on {\bf a.} $[001]$ oriented nanopillars and {\bf b.} $[111]$ oriented nanopillars at varying sizes from 200 nm to 3000 nm, corresponding to the microstructure on cuboidal sides imaged at the maximum load. Proper scale bars were used in different micrographs for a better visualization. \label{fig:ss}}
\end{figure}

The uniaxial compressive stress-strain curves of softer $[001]$ and stiffer $[111]$ oriented pillars at sizes varying from about 200 nm to 3000 nm are presented in Figs. \ref{fig:ss}a and b. Each pillar was loaded until the stress-induced phase transformation from austenite to martensite was completed, then fully unloaded to recover the reverse transformation. More loading curves are shown in the Supplemental Materials, Fig. S4 and S5 \cite{SMref}.
The superelastic behavior, associated with the increasing critical stress of phase transformation, is observed in both orientations. This behavior is evident from the micron scale down to approximately 500 nm. However, when the size falls below 500 nm, the superelasticity begins to weaken in the case of the stiffer [111] oriented pillars, and the residual strain starts to accumulate.
As the critical stress for the martensitic transformation approaches to the flow stress, slips are observed to occur accompanied by phase transformation along [111] compression, captured in 1060 nm, 790 nm, and 270 nm pillars (Fig. \ref{fig:ss}b). Among them, the superelasticity is measurable from the stress-strain curves, characterized as the open hysteresis loop with a partially recoverable plateau strain. The sides of some cuboidal pillars show clear steps due to dislocation slip, which suggests the co-activation of martensitic transformation and plasticity. 

On the other hand, in the softer [001] orientation, superelasticity is preserved down to 200 nm, with gradually increasing critical stresses. The superelastic transformability and reversibility are captured for all tested sizes despite some residual strains. From the SEM images of the pillars at the maximum load, none of the nanopillars show any sign of plastic deformation. The martensitic transformation still dominates the deformation mechanism for the whole range of sizes.  

\begin{figure}
\centering
\includegraphics[width=0.4\textwidth]{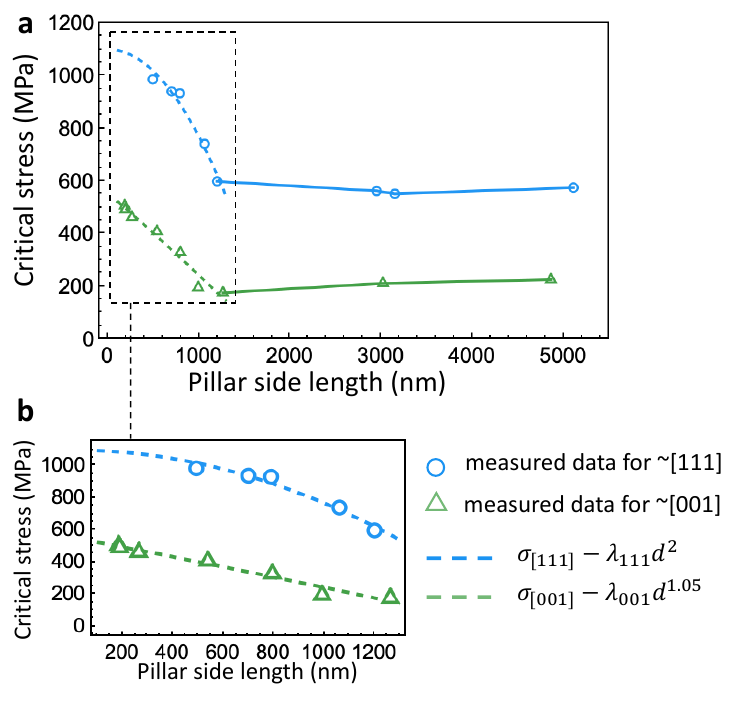}
\caption{{\bf a.} The critical stress of phase transformation varies as the side length of $[001]$ and $[111]$ cuboidal pillars respectively with {\bf b.} submicron region fitted by the new theory. \label{fig:maxload}}
\end{figure}

A quantitative analysis of the critical stress is presented in Fig.\ref{fig:maxload}a, which obviously reveals a bimodal scaling law over the entire size range for both tested orientations. Different scaling laws are separated at the critical size around 1 $\mu$m. In the case of the [001] orientation, micron-sized pillars exhibit a critical stress of approximately 200 MPa, which escalates to 500 MPa for nanopillars of about 200 nm. By contrast, the [111] orientation demonstrates a critical stress of around 600 MPa for micron-sized pillars, which progressively increases and ultimately approaches an asymptote at 1.1 GPa. The stiffer [111] orientation presents a significantly higher critical stress than the softer [001] orientation, accompanied by a distinct scaling trend. This underscores a pronounced size effect dependent on orientation.

Within the blow-up region in Fig. \ref{fig:maxload}b, the scaling of critical stress fits an orientation dependent power decay model as 
\begin{equation}\label{eq:power_law}
    \sigma_c = \sigma_{[uvw]} - \lambda_{[uvw]}{d^n} 
\end{equation}
for the cuboidal pillar side length $0< d < 1000$ nm with positive parameters $\sigma_{[uvw]}$, $\lambda_{[uvw]}$ and $n > 0$. Theoretically, the value of $\sigma_{[uvw]}$ refers to the upper bound of the critical stress as the sample size asymptotically reduces to zero. The geometric interpretation of factor $\lambda_{[uvw]}$ depends on the exponent. By our experiments, the power decay exponent $n=2$ for [111] orientation and $n=1.05$ for [001] orientation. As the critical stress of [111] nanopillars scales quadratically and [001] nanopillars scales linearly, with $\lambda_{[111]} = 0.00033$ as the area scaling factor and $\lambda_{[001]} = 0.215$ as the length scaling factor.

Compared to the yield strength of \CuAlMn alloy: 867MPa for the soft orientation and 1128 MPa for the hard orientation\cite{karami_2022}, it is natural that both martensitic transformation and plasticity can be co-activated in the nanopillars along [111] orientation as $\sigma_{[111]} = 1010$ MPa. When the loading direction is perfectly aligned with the pillar, the martensitic transformation may initiate first. However, it is quickly followed by a slip because the flow stress is sufficiently close to the critical stress required for phase transformation. As a result, we observe slips and martensite twins in some [111] nanopillars (Fig. \ref{fig:ss}b). Additional evidence supporting the combined deformation by slips and martensite transformation can be found in Fig. S4 \cite{SMref}. Even in the case of larger pillars, such as those with a side length of 3 $\mu$m, the slip may occur after the phase transformation, as documented in Supplemental Video S1 \cite{SMref}. At the same length scale (i.e. 3$\mu$m), the in situ observation in Video S2 \cite{SMref} demonstrates that the deformation mechanism can be primarily governed by martensitic transformation, devoid of slips, which was achieved through meticulous control of loading alignment. When it comes to miniature-sized pillars (270 nm), the pillar exhibited slip prior to the phase transformation at stresses exceeding 1 GPa during loading. This in situ deformation was recorded in Supplemental Video S3 \cite{SMref}.

For the [001] nanopillars, the critical stress has an upper limit of $\sigma_{[001]} = 550$ MPa, a value significantly lower than the flow stress. As a result, the martensitic transformation can be detected even in pillars smaller than 200 nm. These experimental findings elucidate the size-dependent deformation mechanisms observed in various types of shape memory alloys. For instance, the martensitic transformation in CuAlNi [001] nanopillars can be characterized due to the critical stress limit being substantially lower than its flow stress. Conversely, in NiTi nanopillars, the critical stress is inherently much larger. As the size diminishes, the critical stress of NiTi readily approaches the flow stress. Consequently, the pillar may yield prior to phase transformation due to stress localization during the loading process.

\begin{figure*}
\centering
\includegraphics[width=0.7\textwidth]{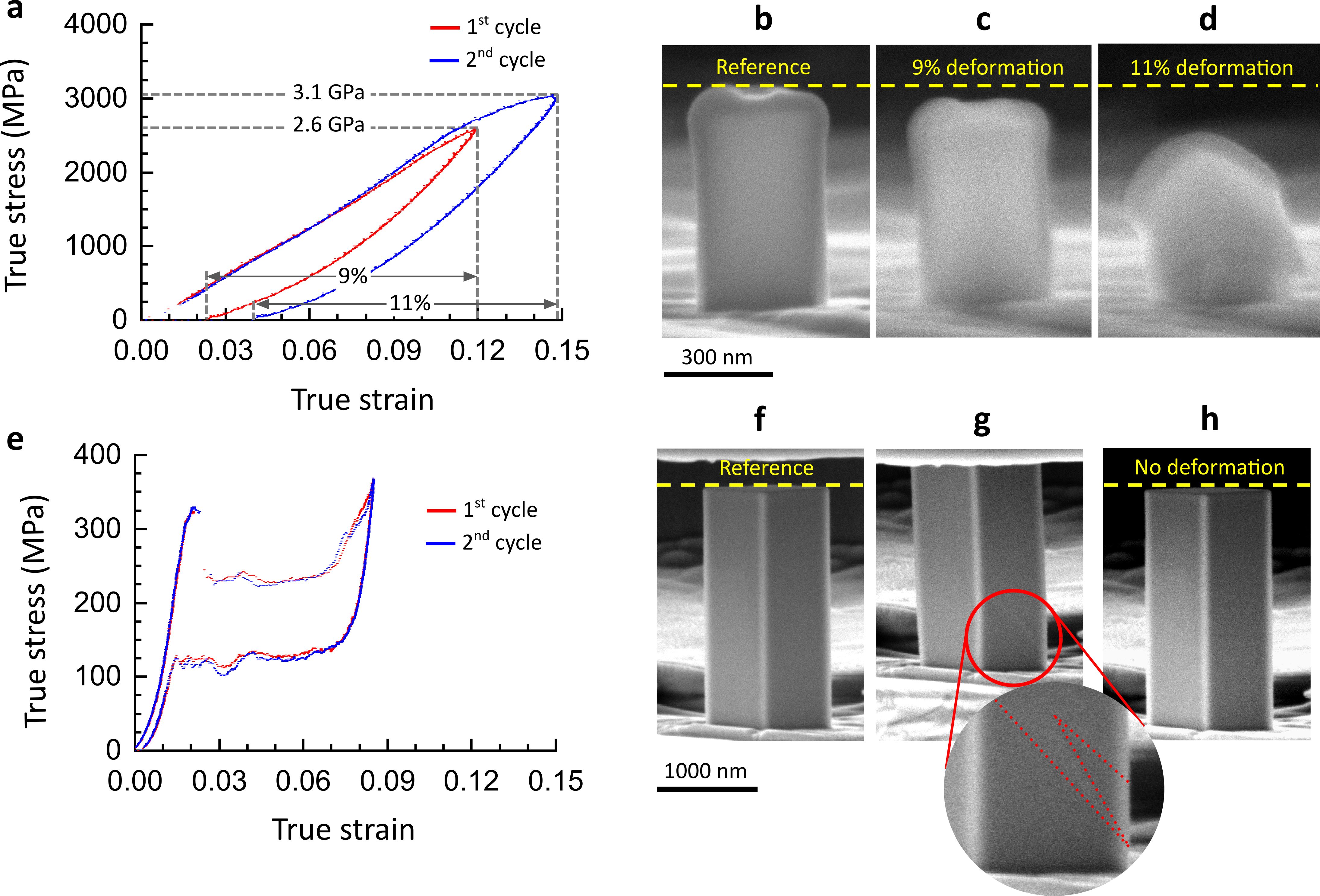}
\caption{{\bf a.} The stress and strain curves of the [001] oriented nanopillar with side length of 220 nm under two consecutive uniaxial nanocompression cycles, corresponding to {\bf b.} the reference, {\bf c.} 9\% deformed and {\bf d.} 11\% deformed configurations. {\bf e.} The stress-strain curve of the [001] oriented nanopillar with side length of 896 nm under two consecutive compression cycles, corresponding to {\bf f.} the reference configuration, {\bf g.} the deformed configuration at maximum compressive stress, and {\bf h.} the undeformed configuration after unloading. The inset indicates the formation of martensite twins. \label{fig:tiny}}
\end{figure*}

The theoretical scaling model in \eqref{eq:power_law} for the soft orientation [001] suggests a big potential for achieving high transformability under high stress. We select several nanopillars around 200 nm to investigate the ultimate elasticity that can be achieved by further increasing the maximum load. Fig \ref{fig:tiny}a shows the stress-strain curves of a 220 nm cuboidal pillar under the uniaxial nanocompression in two consecutive loading cycles. During the first cycle, the nanopillar surprisingly withstands up to 2.6 GPa while recovering 9\% strain upon unloading. The strength and elastic deformation become more pronounced during the second cycle. The nanopillar recovers 11\% strain from 3.1 GPa compressive stress. The reference configuration as well as the deformed configurations after two magnificent stress cycles are shown in Figs. \ref{fig:tiny}b-d. Approximately 2\% and 4\% residual strains were accumulated after the first and second cycles respectively. The high transformability is achieved at a high stress over 2 GPa in 220 nm pillar without showing noticeable superelastic plateau. In contrast, Fig. \ref{fig:tiny}e exhibits the normal superelastic behavior captured in a 890 nm cuboidal pillar that forms normal martensite microstructures during loading, as seen in Figs. \ref{fig:tiny}f-h.      

The deformation mechanisms, as seen in Supplemental Videos S4 and S5 \cite{SMref} for two other nanopillars measuring 205 nm and 185 nm, are sensitive to both the loading condition and alignment. The 205 nm pillar exhibited a high recoverable strain at a high stress level (approximately 2 GPa), while the 185 nm pillar demonstrated a superelastic plateau strain at around 500 MPa. More loading curves of nanopillars around 200 nm were given in Supplemental Materials, Fig. S5 \cite{SMref}.

To explore the deformation mechanism at the nanoscale for the soft orientation [001], we utilized high-resolution transmission electron microscopy (HRTEM) to examine the local lattice distortion of the post-deformed nanopillars with approximately 200 nm in size. For comparison, we also examined the 890 nm pillar, which exhibits normal superelasticity under moderate stress.

\begin{figure*}
\centering
\includegraphics[width=0.6\textwidth]{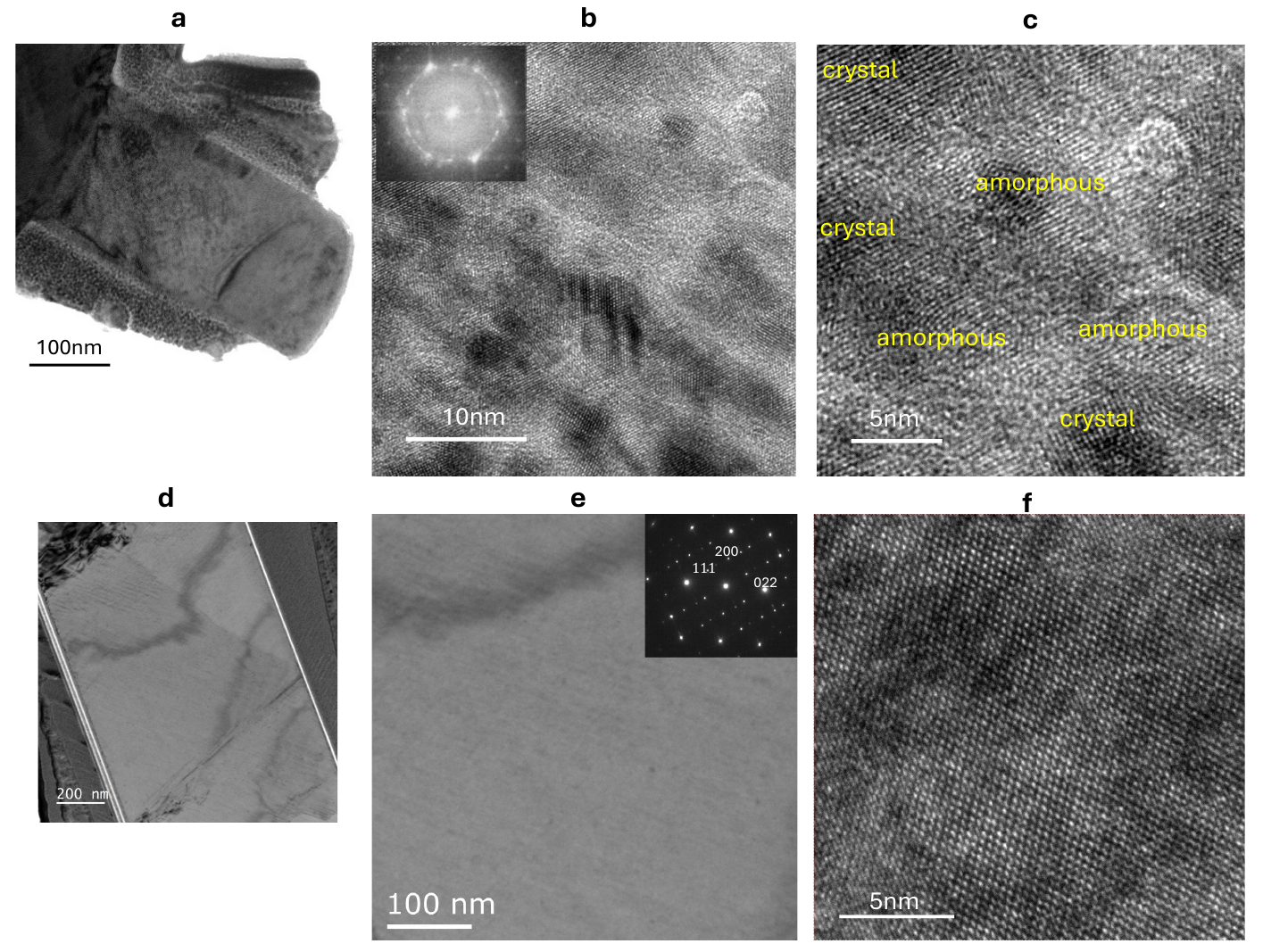}
\caption{The TEM studies of nominal [001] nanopillar at size 220 nm {\bf a}-{\bf c} and 890 nm {\bf d} - {\bf f} after removal of stress. 
{\bf a.} Low magnification bright field image of the 220 nm pillar, corresponding to {\bf b.} and {\bf c.} High-resolution TEM (HRTEM) images showing nanoscale amorphous clusters. The Fourier Transform of the HRTEM is shown as the inset. Low magnification bright field images of the 890 nm pillar are shown in {\bf d.} and {\bf e.}. The inset of {\bf e} is the selected area diffraction pattern with zone axis along $[01\bar 1]$. {\bf f.} HRTEM image showing a pure crystalline structure of 890 nm pillar. \label{fig:TEM}}
\end{figure*}

Figs. \ref{fig:TEM}a and d provide a low magnification view of the microstructure. The central portion of each pillar was magnified under TEM, as shown in Fig. \ref{fig:TEM}b for the 200 nm nanopillar and Fig. \ref{fig:TEM}e for the 890 nm pillar. In the 200 nm nanopillar, amorphization was identified in the high-resolution TEM images. In contrast, the selected area electron diffraction in Figure \ref{fig:TEM}e indicates that the 890 nm pillar retains its crystalline status. The diffraction pattern can be indexed by the space group $Fm\Bar{3}m$. 

Fig. \ref{fig:TEM}c demonstrates that local lattice frustration leads to the emergence of amorphous clusters within the 200 nm nanopillar. These clusters, which are amorphous in nature, are embedded in the crystalline matrix. The amorphous-crystalline heterogeneous structure has the ability to relax elastically even after significant stress levels. This characteristic is reminiscent of the behavior observed in nanocrystalline nitinol alloys \cite{hua_2018, hua_2021nat}. A distinguishing feature here is the initiation of the amorphization process by the single crystal lattice frustration. Notably, the size of this nanopillar significantly exceeds the grain sizes commonly reported in the literature. Such an interaction expands the potential applications of Cu-based shape memory alloys, particularly in the realm of miniaturized devices.

To summarize, we conducted an in situ nanomechanical compression experiment to explore the size effect and scaling law of \CuAlMn in two primary orientations: the nominal stiff direction [111] and the nominal soft direction [001]. Our findings revealed that the size effect on the critical stress for the martensitic phase transformation does not satisfy a universal power law from micron to nanometer sizes. An orientation-dependent power decay law suggests a specific increase in critical stress for pillars smaller than 1 micron meter. Most notably, we discovered high deformability with exceptional recovery capability under high stress (2 GPa) along the elastically soft direction [001] when the nanopillar sizes are reduced to 200 nm. This research paves the way for revealing the superior superelastic behavior of shape memory alloys for nanotechnology applications.

\begin{acknowledgments}
M. K and X.C. thank the financial support under GRF Grants No. 16203021, 16204022 and No. 16203023 by Research Grants Council, Hong Kong.
\end{acknowledgments}

\bibliography{sn-bibliography}

\begin{thebibliography}{31}%
\makeatletter
\providecommand \@ifxundefined [1]{%
 \@ifx{#1\undefined}
}%
\providecommand \@ifnum [1]{%
 \ifnum #1\expandafter \@firstoftwo
 \else \expandafter \@secondoftwo
 \fi
}%
\providecommand \@ifx [1]{%
 \ifx #1\expandafter \@firstoftwo
 \else \expandafter \@secondoftwo
 \fi
}%
\providecommand \natexlab [1]{#1}%
\providecommand \enquote  [1]{``#1''}%
\providecommand \bibnamefont  [1]{#1}%
\providecommand \bibfnamefont [1]{#1}%
\providecommand \citenamefont [1]{#1}%
\providecommand \href@noop [0]{\@secondoftwo}%
\providecommand \href [0]{\begingroup \@sanitize@url \@href}%
\providecommand \@href[1]{\@@startlink{#1}\@@href}%
\providecommand \@@href[1]{\endgroup#1\@@endlink}%
\providecommand \@sanitize@url [0]{\catcode `\\12\catcode `\$12\catcode
  `\&12\catcode `\#12\catcode `\^12\catcode `\_12\catcode `\%12\relax}%
\providecommand \@@startlink[1]{}%
\providecommand \@@endlink[0]{}%
\providecommand \url  [0]{\begingroup\@sanitize@url \@url }%
\providecommand \@url [1]{\endgroup\@href {#1}{\urlprefix }}%
\providecommand \urlprefix  [0]{URL }%
\providecommand \Eprint [0]{\href }%
\providecommand \doibase [0]{https://doi.org/}%
\providecommand \selectlanguage [0]{\@gobble}%
\providecommand \bibinfo  [0]{\@secondoftwo}%
\providecommand \bibfield  [0]{\@secondoftwo}%
\providecommand \translation [1]{[#1]}%
\providecommand \BibitemOpen [0]{}%
\providecommand \bibitemStop [0]{}%
\providecommand \bibitemNoStop [0]{.\EOS\space}%
\providecommand \EOS [0]{\spacefactor3000\relax}%
\providecommand \BibitemShut  [1]{\csname bibitem#1\endcsname}%
\let\auto@bib@innerbib\@empty
\bibitem [{\citenamefont {Levi}\ \emph {et~al.}(2008)\citenamefont {Levi},
  \citenamefont {Kusnezov},\ and\ \citenamefont {Carman}}]{levi_2008}%
  \BibitemOpen
  \bibfield  {author} {\bibinfo {author} {\bibfnamefont {D.~S.}\ \bibnamefont
  {Levi}}, \bibinfo {author} {\bibfnamefont {N.}~\bibnamefont {Kusnezov}},\
  and\ \bibinfo {author} {\bibfnamefont {G.~P.}\ \bibnamefont {Carman}},\
  }\bibfield  {title} {\bibinfo {title} {Smart materials applications for
  pediatric cardiovascular devices},\ }\href@noop {} {\bibfield  {journal}
  {\bibinfo  {journal} {Pediatric research}\ }\textbf {\bibinfo {volume}
  {63}},\ \bibinfo {pages} {552} (\bibinfo {year} {2008})}\BibitemShut
  {NoStop}%
\bibitem [{\citenamefont {Jansen}\ \emph {et~al.}(1992)\citenamefont {Jansen},
  \citenamefont {Willeke}, \citenamefont {Reul},\ and\ \citenamefont
  {RUM}}]{jansen_1992}%
  \BibitemOpen
  \bibfield  {author} {\bibinfo {author} {\bibfnamefont {J.}~\bibnamefont
  {Jansen}}, \bibinfo {author} {\bibfnamefont {S.}~\bibnamefont {Willeke}},
  \bibinfo {author} {\bibfnamefont {H.}~\bibnamefont {Reul}},\ and\ \bibinfo
  {author} {\bibfnamefont {G.}~\bibnamefont {RUM}},\ }\bibfield  {title}
  {\bibinfo {title} {Detachable shape-memory sewing ring for heart valves},\
  }\href@noop {} {\bibfield  {journal} {\bibinfo  {journal} {Artificial
  Organs}\ }\textbf {\bibinfo {volume} {16}},\ \bibinfo {pages} {294} (\bibinfo
  {year} {1992})}\BibitemShut {NoStop}%
\bibitem [{\citenamefont {Chen}\ \emph {et~al.}(2007)\citenamefont {Chen},
  \citenamefont {Tsai}, \citenamefont {Chang}, \citenamefont {Lai},
  \citenamefont {Mi}, \citenamefont {Liu}, \citenamefont {Wong},\ and\
  \citenamefont {Sung}}]{chen_2007rapidly}%
  \BibitemOpen
  \bibfield  {author} {\bibinfo {author} {\bibfnamefont {M.-C.}\ \bibnamefont
  {Chen}}, \bibinfo {author} {\bibfnamefont {H.-W.}\ \bibnamefont {Tsai}},
  \bibinfo {author} {\bibfnamefont {Y.}~\bibnamefont {Chang}}, \bibinfo
  {author} {\bibfnamefont {W.-Y.}\ \bibnamefont {Lai}}, \bibinfo {author}
  {\bibfnamefont {F.-L.}\ \bibnamefont {Mi}}, \bibinfo {author} {\bibfnamefont
  {C.-T.}\ \bibnamefont {Liu}}, \bibinfo {author} {\bibfnamefont {H.-S.}\
  \bibnamefont {Wong}},\ and\ \bibinfo {author} {\bibfnamefont {H.-W.}\
  \bibnamefont {Sung}},\ }\bibfield  {title} {\bibinfo {title} {Rapidly
  self-expandable polymeric stents with a shape-memory property},\ }\href@noop
  {} {\bibfield  {journal} {\bibinfo  {journal} {Biomacromolecules}\ }\textbf
  {\bibinfo {volume} {8}},\ \bibinfo {pages} {2774} (\bibinfo {year}
  {2007})}\BibitemShut {NoStop}%
\bibitem [{\citenamefont {Yeazel}\ and\ \citenamefont
  {Becker}(2020)}]{yeazel_2020}%
  \BibitemOpen
  \bibfield  {author} {\bibinfo {author} {\bibfnamefont {T.~R.}\ \bibnamefont
  {Yeazel}}\ and\ \bibinfo {author} {\bibfnamefont {M.~L.}\ \bibnamefont
  {Becker}},\ }\bibfield  {title} {\bibinfo {title} {Advancing toward 3d
  printing of bioresorbable shape memory polymer stents},\ }\href@noop {}
  {\bibfield  {journal} {\bibinfo  {journal} {Biomacromolecules}\ }\textbf
  {\bibinfo {volume} {21}},\ \bibinfo {pages} {3957} (\bibinfo {year}
  {2020})}\BibitemShut {NoStop}%
\bibitem [{\citenamefont {Bahraminasab}\ and\ \citenamefont
  {Sahari}(2013)}]{bahraminasab_2013}%
  \BibitemOpen
  \bibfield  {author} {\bibinfo {author} {\bibfnamefont {M.}~\bibnamefont
  {Bahraminasab}}\ and\ \bibinfo {author} {\bibfnamefont {B.~B.}\ \bibnamefont
  {Sahari}},\ }\bibfield  {title} {\bibinfo {title} {Niti shape memory alloys,
  promising materials in orthopedic applications},\ }\href@noop {} {\bibfield
  {journal} {\bibinfo  {journal} {Shape Memory Alloys-Processing,
  Characterization and Applications}\ ,\ \bibinfo {pages} {261}} (\bibinfo
  {year} {2013})}\BibitemShut {NoStop}%
\bibitem [{\citenamefont {Bechtold}\ \emph
  {et~al.}(2016{\natexlab{a}})\citenamefont {Bechtold}, \citenamefont {Lima~de
  Miranda}, \citenamefont {Chluba}, \citenamefont {Zamponi},\ and\
  \citenamefont {Quandt}}]{bechtold_2016method}%
  \BibitemOpen
  \bibfield  {author} {\bibinfo {author} {\bibfnamefont {C.}~\bibnamefont
  {Bechtold}}, \bibinfo {author} {\bibfnamefont {R.}~\bibnamefont {Lima~de
  Miranda}}, \bibinfo {author} {\bibfnamefont {C.}~\bibnamefont {Chluba}},
  \bibinfo {author} {\bibfnamefont {C.}~\bibnamefont {Zamponi}},\ and\ \bibinfo
  {author} {\bibfnamefont {E.}~\bibnamefont {Quandt}},\ }\bibfield  {title}
  {\bibinfo {title} {Method for fabricating miniaturized {NiTi} self-expandable
  thin film devices with increased radiopacity},\ }\href@noop {} {\bibfield
  {journal} {\bibinfo  {journal} {Shape Memory and Superelasticity}\ }\textbf
  {\bibinfo {volume} {2}},\ \bibinfo {pages} {391} (\bibinfo {year}
  {2016}{\natexlab{a}})}\BibitemShut {NoStop}%
\bibitem [{\citenamefont {Bechtold}\ \emph
  {et~al.}(2016{\natexlab{b}})\citenamefont {Bechtold}, \citenamefont
  {de~Miranda}, \citenamefont {Chluba},\ and\ \citenamefont
  {Quandt}}]{bechtold2016fab}%
  \BibitemOpen
  \bibfield  {author} {\bibinfo {author} {\bibfnamefont {C.}~\bibnamefont
  {Bechtold}}, \bibinfo {author} {\bibfnamefont {R.~L.}\ \bibnamefont
  {de~Miranda}}, \bibinfo {author} {\bibfnamefont {C.}~\bibnamefont {Chluba}},\
  and\ \bibinfo {author} {\bibfnamefont {E.}~\bibnamefont {Quandt}},\
  }\bibfield  {title} {\bibinfo {title} {Fabrication of self-expandable {NiTi}
  thin film devices with micro-electrode array for bioelectric sensing,
  stimulation and ablation},\ }\href@noop {} {\bibfield  {journal} {\bibinfo
  {journal} {Biomedical microdevices}\ }\textbf {\bibinfo {volume} {18}},\
  \bibinfo {pages} {106} (\bibinfo {year} {2016}{\natexlab{b}})}\BibitemShut
  {NoStop}%
\bibitem [{\citenamefont {Frick}\ \emph {et~al.}(2007)\citenamefont {Frick},
  \citenamefont {Orso},\ and\ \citenamefont {Arzt}}]{frick_2007}%
  \BibitemOpen
  \bibfield  {author} {\bibinfo {author} {\bibfnamefont {C.}~\bibnamefont
  {Frick}}, \bibinfo {author} {\bibfnamefont {S.}~\bibnamefont {Orso}},\ and\
  \bibinfo {author} {\bibfnamefont {E.}~\bibnamefont {Arzt}},\ }\bibfield
  {title} {\bibinfo {title} {Loss of pseudoelasticity in nickel--titanium
  sub-micron compression pillars},\ }\href@noop {} {\bibfield  {journal}
  {\bibinfo  {journal} {Acta Materialia}\ }\textbf {\bibinfo {volume} {55}},\
  \bibinfo {pages} {3845} (\bibinfo {year} {2007})}\BibitemShut {NoStop}%
\bibitem [{\citenamefont {Ozdemir}\ \emph {et~al.}(2012)\citenamefont
  {Ozdemir}, \citenamefont {Karaman}, \citenamefont {Mara}, \citenamefont
  {Chumlyakov},\ and\ \citenamefont {Karaca}}]{ozdemir2012size}%
  \BibitemOpen
  \bibfield  {author} {\bibinfo {author} {\bibfnamefont {N.}~\bibnamefont
  {Ozdemir}}, \bibinfo {author} {\bibfnamefont {I.}~\bibnamefont {Karaman}},
  \bibinfo {author} {\bibfnamefont {N.}~\bibnamefont {Mara}}, \bibinfo {author}
  {\bibfnamefont {Y.~I.}\ \bibnamefont {Chumlyakov}},\ and\ \bibinfo {author}
  {\bibfnamefont {H.}~\bibnamefont {Karaca}},\ }\bibfield  {title} {\bibinfo
  {title} {Size effects in the superelastic response of {Ni54Fe19Ga27} shape
  memory alloy pillars with a two stage martensitic transformation},\
  }\href@noop {} {\bibfield  {journal} {\bibinfo  {journal} {Acta Materialia}\
  }\textbf {\bibinfo {volume} {60}},\ \bibinfo {pages} {5670} (\bibinfo {year}
  {2012})}\BibitemShut {NoStop}%
\bibitem [{\citenamefont {Gomez-Cortes}\ \emph {et~al.}(2017)\citenamefont
  {Gomez-Cortes}, \citenamefont {No}, \citenamefont {Lopez-Ferreno},
  \citenamefont {Hernandez-Saz}, \citenamefont {Molina}, \citenamefont
  {Chuvilin},\ and\ \citenamefont {San~Juan}}]{gomez_2017}%
  \BibitemOpen
  \bibfield  {author} {\bibinfo {author} {\bibfnamefont {J.~F.}\ \bibnamefont
  {Gomez-Cortes}}, \bibinfo {author} {\bibfnamefont {M.~L.}\ \bibnamefont
  {No}}, \bibinfo {author} {\bibfnamefont {I.}~\bibnamefont {Lopez-Ferreno}},
  \bibinfo {author} {\bibfnamefont {J.}~\bibnamefont {Hernandez-Saz}}, \bibinfo
  {author} {\bibfnamefont {S.~I.}\ \bibnamefont {Molina}}, \bibinfo {author}
  {\bibfnamefont {A.}~\bibnamefont {Chuvilin}},\ and\ \bibinfo {author}
  {\bibfnamefont {J.~M.}\ \bibnamefont {San~Juan}},\ }\bibfield  {title}
  {\bibinfo {title} {Size effect and scaling power-law for superelasticity in
  shape-memory alloys at the nanoscale},\ }\href@noop {} {\bibfield  {journal}
  {\bibinfo  {journal} {Nature nanotechnology}\ }\textbf {\bibinfo {volume}
  {12}},\ \bibinfo {pages} {790} (\bibinfo {year} {2017})}\BibitemShut
  {NoStop}%
\bibitem [{\citenamefont {Frick}\ \emph {et~al.}(2010)\citenamefont {Frick},
  \citenamefont {Clark}, \citenamefont {Schneider}, \citenamefont {Maa{\ss}},
  \citenamefont {Van~Petegem},\ and\ \citenamefont
  {Van~Swygenhoven}}]{frick_2010}%
  \BibitemOpen
  \bibfield  {author} {\bibinfo {author} {\bibfnamefont {C.~P.}\ \bibnamefont
  {Frick}}, \bibinfo {author} {\bibfnamefont {B.~G.}\ \bibnamefont {Clark}},
  \bibinfo {author} {\bibfnamefont {A.~S.}\ \bibnamefont {Schneider}}, \bibinfo
  {author} {\bibfnamefont {R.}~\bibnamefont {Maa{\ss}}}, \bibinfo {author}
  {\bibfnamefont {S.}~\bibnamefont {Van~Petegem}},\ and\ \bibinfo {author}
  {\bibfnamefont {H.}~\bibnamefont {Van~Swygenhoven}},\ }\bibfield  {title}
  {\bibinfo {title} {On the plasticity of small-scale nickel--titanium shape
  memory alloys},\ }\href@noop {} {\bibfield  {journal} {\bibinfo  {journal}
  {Scripta Materialia}\ }\textbf {\bibinfo {volume} {62}},\ \bibinfo {pages}
  {492} (\bibinfo {year} {2010})}\BibitemShut {NoStop}%
\bibitem [{\citenamefont {Ahadi}\ and\ \citenamefont
  {Sun}(2014)}]{ahadi2014effects}%
  \BibitemOpen
  \bibfield  {author} {\bibinfo {author} {\bibfnamefont {A.}~\bibnamefont
  {Ahadi}}\ and\ \bibinfo {author} {\bibfnamefont {Q.}~\bibnamefont {Sun}},\
  }\bibfield  {title} {\bibinfo {title} {Effects of grain size on the
  rate-dependent thermomechanical responses of nanostructured superelastic
  {NiTi}},\ }\href@noop {} {\bibfield  {journal} {\bibinfo  {journal} {Acta
  Materialia}\ }\textbf {\bibinfo {volume} {76}},\ \bibinfo {pages} {186}
  (\bibinfo {year} {2014})}\BibitemShut {NoStop}%
\bibitem [{\citenamefont {Xiao}\ \emph {et~al.}(2017)\citenamefont {Xiao},
  \citenamefont {Zeng},\ and\ \citenamefont {Lei}}]{xiao2017grain}%
  \BibitemOpen
  \bibfield  {author} {\bibinfo {author} {\bibfnamefont {Y.}~\bibnamefont
  {Xiao}}, \bibinfo {author} {\bibfnamefont {P.}~\bibnamefont {Zeng}},\ and\
  \bibinfo {author} {\bibfnamefont {L.}~\bibnamefont {Lei}},\ }\bibfield
  {title} {\bibinfo {title} {Grain size effect on mechanical performance of
  nanostructured superelastic niti alloy},\ }\href@noop {} {\bibfield
  {journal} {\bibinfo  {journal} {Materials Research Express}\ }\textbf
  {\bibinfo {volume} {4}},\ \bibinfo {pages} {035702} (\bibinfo {year}
  {2017})}\BibitemShut {NoStop}%
\bibitem [{\citenamefont {Hua}\ \emph {et~al.}(2018)\citenamefont {Hua},
  \citenamefont {Chu},\ and\ \citenamefont {Sun}}]{hua_2018}%
  \BibitemOpen
  \bibfield  {author} {\bibinfo {author} {\bibfnamefont {P.}~\bibnamefont
  {Hua}}, \bibinfo {author} {\bibfnamefont {K.}~\bibnamefont {Chu}},\ and\
  \bibinfo {author} {\bibfnamefont {Q.}~\bibnamefont {Sun}},\ }\bibfield
  {title} {\bibinfo {title} {Grain refinement and amorphization in
  nanocrystalline niti micropillars under uniaxial compression},\ }\href@noop
  {} {\bibfield  {journal} {\bibinfo  {journal} {Scripta Materialia}\ }\textbf
  {\bibinfo {volume} {154}},\ \bibinfo {pages} {123} (\bibinfo {year}
  {2018})}\BibitemShut {NoStop}%
\bibitem [{\citenamefont {Hua}\ \emph {et~al.}(2021)\citenamefont {Hua},
  \citenamefont {Xia}, \citenamefont {Onuki},\ and\ \citenamefont
  {Sun}}]{hua_2021nat}%
  \BibitemOpen
  \bibfield  {author} {\bibinfo {author} {\bibfnamefont {P.}~\bibnamefont
  {Hua}}, \bibinfo {author} {\bibfnamefont {M.}~\bibnamefont {Xia}}, \bibinfo
  {author} {\bibfnamefont {Y.}~\bibnamefont {Onuki}},\ and\ \bibinfo {author}
  {\bibfnamefont {Q.}~\bibnamefont {Sun}},\ }\bibfield  {title} {\bibinfo
  {title} {Nanocomposite niti shape memory alloy with high strength and fatigue
  resistance},\ }\href@noop {} {\bibfield  {journal} {\bibinfo  {journal}
  {Nature Nanotechnology}\ }\textbf {\bibinfo {volume} {16}},\ \bibinfo {pages}
  {409} (\bibinfo {year} {2021})}\BibitemShut {NoStop}%
\bibitem [{\citenamefont {Juan}\ \emph {et~al.}(2009)\citenamefont {Juan},
  \citenamefont {N{\'o}},\ and\ \citenamefont {Schuh}}]{juan_2009nano}%
  \BibitemOpen
  \bibfield  {author} {\bibinfo {author} {\bibfnamefont {J.~S.}\ \bibnamefont
  {Juan}}, \bibinfo {author} {\bibfnamefont {M.~L.}\ \bibnamefont {N{\'o}}},\
  and\ \bibinfo {author} {\bibfnamefont {C.~A.}\ \bibnamefont {Schuh}},\
  }\bibfield  {title} {\bibinfo {title} {Nanoscale shape-memory alloys for
  ultrahigh mechanical damping},\ }\href@noop {} {\bibfield  {journal}
  {\bibinfo  {journal} {Nature nanotechnology}\ }\textbf {\bibinfo {volume}
  {4}},\ \bibinfo {pages} {415} (\bibinfo {year} {2009})}\BibitemShut {NoStop}%
\bibitem [{\citenamefont {San~Juan}\ and\ \citenamefont
  {N{\'o}}(2013)}]{san2013superelasticity}%
  \BibitemOpen
  \bibfield  {author} {\bibinfo {author} {\bibfnamefont {J.}~\bibnamefont
  {San~Juan}}\ and\ \bibinfo {author} {\bibfnamefont {M.}~\bibnamefont
  {N{\'o}}},\ }\bibfield  {title} {\bibinfo {title} {Superelasticity and shape
  memory at nano-scale: Size effects on the martensitic transformation},\
  }\href@noop {} {\bibfield  {journal} {\bibinfo  {journal} {Journal of Alloys
  and Compounds}\ }\textbf {\bibinfo {volume} {577}},\ \bibinfo {pages} {S25}
  (\bibinfo {year} {2013})}\BibitemShut {NoStop}%
\bibitem [{\citenamefont {Fuster}\ \emph {et~al.}(2020)\citenamefont {Fuster},
  \citenamefont {G{\'o}mez-Cort{\'e}s}, \citenamefont {N{\'o}},\ and\
  \citenamefont {San~Juan}}]{fuster_2020}%
  \BibitemOpen
  \bibfield  {author} {\bibinfo {author} {\bibfnamefont {V.}~\bibnamefont
  {Fuster}}, \bibinfo {author} {\bibfnamefont {J.~F.}\ \bibnamefont
  {G{\'o}mez-Cort{\'e}s}}, \bibinfo {author} {\bibfnamefont {M.~L.}\
  \bibnamefont {N{\'o}}},\ and\ \bibinfo {author} {\bibfnamefont {J.~M.}\
  \bibnamefont {San~Juan}},\ }\bibfield  {title} {\bibinfo {title} {Universal
  scaling law for the size effect on superelasticity at the nanoscale promotes
  the use of shape-memory alloys in stretchable devices},\ }\href@noop {}
  {\bibfield  {journal} {\bibinfo  {journal} {Advanced Electronic Materials}\
  }\textbf {\bibinfo {volume} {6}},\ \bibinfo {pages} {1900741} (\bibinfo
  {year} {2020})}\BibitemShut {NoStop}%
\bibitem [{\citenamefont {Dou}\ and\ \citenamefont
  {Derby}(2009)}]{dou2009universal}%
  \BibitemOpen
  \bibfield  {author} {\bibinfo {author} {\bibfnamefont {R.}~\bibnamefont
  {Dou}}\ and\ \bibinfo {author} {\bibfnamefont {B.}~\bibnamefont {Derby}},\
  }\bibfield  {title} {\bibinfo {title} {A universal scaling law for the
  strength of metal micropillars and nanowires},\ }\href@noop {} {\bibfield
  {journal} {\bibinfo  {journal} {Scripta Materialia}\ }\textbf {\bibinfo
  {volume} {61}},\ \bibinfo {pages} {524} (\bibinfo {year} {2009})}\BibitemShut
  {NoStop}%
\bibitem [{\citenamefont {Jennings}\ and\ \citenamefont
  {Greer}(2011)}]{jennings_2011}%
  \BibitemOpen
  \bibfield  {author} {\bibinfo {author} {\bibfnamefont {A.~T.}\ \bibnamefont
  {Jennings}}\ and\ \bibinfo {author} {\bibfnamefont {J.~R.}\ \bibnamefont
  {Greer}},\ }\bibfield  {title} {\bibinfo {title} {Tensile deformation of
  electroplated copper nanopillars},\ }\href
  {https://doi.org/10.1080/14786435.2010.505180} {\bibfield  {journal}
  {\bibinfo  {journal} {Philosophical Magazine}\ }\textbf {\bibinfo {volume}
  {91}},\ \bibinfo {pages} {1108} (\bibinfo {year} {2011})}\BibitemShut
  {NoStop}%
\bibitem [{\citenamefont {Greer}\ \emph {et~al.}(2005)\citenamefont {Greer},
  \citenamefont {Oliver},\ and\ \citenamefont {Nix}}]{greer_2005}%
  \BibitemOpen
  \bibfield  {author} {\bibinfo {author} {\bibfnamefont {J.~R.}\ \bibnamefont
  {Greer}}, \bibinfo {author} {\bibfnamefont {W.~C.}\ \bibnamefont {Oliver}},\
  and\ \bibinfo {author} {\bibfnamefont {W.~D.}\ \bibnamefont {Nix}},\
  }\bibfield  {title} {\bibinfo {title} {Size dependence of mechanical
  properties of gold at the micron scale in the absence of strain gradients},\
  }\href@noop {} {\bibfield  {journal} {\bibinfo  {journal} {Acta Materialia}\
  }\textbf {\bibinfo {volume} {53}},\ \bibinfo {pages} {1821} (\bibinfo {year}
  {2005})}\BibitemShut {NoStop}%
\bibitem [{\citenamefont {Volkert}\ and\ \citenamefont
  {Lilleodden}(2006)}]{volkert_2006}%
  \BibitemOpen
  \bibfield  {author} {\bibinfo {author} {\bibfnamefont {C.~A.}\ \bibnamefont
  {Volkert}}\ and\ \bibinfo {author} {\bibfnamefont {E.~T.}\ \bibnamefont
  {Lilleodden}},\ }\bibfield  {title} {\bibinfo {title} {Size effects in the
  deformation of sub-micron au columns},\ }\href@noop {} {\bibfield  {journal}
  {\bibinfo  {journal} {Philosophical Magazine}\ }\textbf {\bibinfo {volume}
  {86}},\ \bibinfo {pages} {5567} (\bibinfo {year} {2006})}\BibitemShut
  {NoStop}%
\bibitem [{\citenamefont {Dimiduk}\ \emph {et~al.}(2005)\citenamefont
  {Dimiduk}, \citenamefont {Uchic},\ and\ \citenamefont
  {Parthasarathy}}]{dimiduk_2005}%
  \BibitemOpen
  \bibfield  {author} {\bibinfo {author} {\bibfnamefont {D.}~\bibnamefont
  {Dimiduk}}, \bibinfo {author} {\bibfnamefont {M.}~\bibnamefont {Uchic}},\
  and\ \bibinfo {author} {\bibfnamefont {T.}~\bibnamefont {Parthasarathy}},\
  }\bibfield  {title} {\bibinfo {title} {Size-affected single-slip behavior of
  pure nickel microcrystals},\ }\href@noop {} {\bibfield  {journal} {\bibinfo
  {journal} {Acta materialia}\ }\textbf {\bibinfo {volume} {53}},\ \bibinfo
  {pages} {4065} (\bibinfo {year} {2005})}\BibitemShut {NoStop}%
\bibitem [{\citenamefont {Kim}\ \emph {et~al.}(2010)\citenamefont {Kim},
  \citenamefont {Jang},\ and\ \citenamefont {Greer}}]{kim_2010}%
  \BibitemOpen
  \bibfield  {author} {\bibinfo {author} {\bibfnamefont {J.-Y.}\ \bibnamefont
  {Kim}}, \bibinfo {author} {\bibfnamefont {D.}~\bibnamefont {Jang}},\ and\
  \bibinfo {author} {\bibfnamefont {J.~R.}\ \bibnamefont {Greer}},\ }\bibfield
  {title} {\bibinfo {title} {Tensile and compressive behavior of tungsten,
  molybdenum, tantalum and niobium at the nanoscale},\ }\href@noop {}
  {\bibfield  {journal} {\bibinfo  {journal} {Acta Materialia}\ }\textbf
  {\bibinfo {volume} {58}},\ \bibinfo {pages} {2355} (\bibinfo {year}
  {2010})}\BibitemShut {NoStop}%
\bibitem [{\citenamefont {Greer}\ and\ \citenamefont
  {De~Hosson}(2011)}]{greer_2011rev}%
  \BibitemOpen
  \bibfield  {author} {\bibinfo {author} {\bibfnamefont {J.~R.}\ \bibnamefont
  {Greer}}\ and\ \bibinfo {author} {\bibfnamefont {J.~T.~M.}\ \bibnamefont
  {De~Hosson}},\ }\bibfield  {title} {\bibinfo {title} {Plasticity in
  small-sized metallic systems: Intrinsic versus extrinsic size effect},\
  }\href@noop {} {\bibfield  {journal} {\bibinfo  {journal} {Progress in
  Materials Science}\ }\textbf {\bibinfo {volume} {56}},\ \bibinfo {pages}
  {654} (\bibinfo {year} {2011})}\BibitemShut {NoStop}%
\bibitem [{\citenamefont {Kim}\ \emph {et~al.}(2012)\citenamefont {Kim},
  \citenamefont {Jang},\ and\ \citenamefont {Greer}}]{kim_2012}%
  \BibitemOpen
  \bibfield  {author} {\bibinfo {author} {\bibfnamefont {J.-Y.}\ \bibnamefont
  {Kim}}, \bibinfo {author} {\bibfnamefont {D.}~\bibnamefont {Jang}},\ and\
  \bibinfo {author} {\bibfnamefont {J.~R.}\ \bibnamefont {Greer}},\ }\bibfield
  {title} {\bibinfo {title} {Crystallographic orientation and size dependence
  of tension--compression asymmetry in molybdenum nano-pillars},\ }\href@noop
  {} {\bibfield  {journal} {\bibinfo  {journal} {International Journal of
  Plasticity}\ }\textbf {\bibinfo {volume} {28}},\ \bibinfo {pages} {46}
  (\bibinfo {year} {2012})}\BibitemShut {NoStop}%
\bibitem [{\citenamefont {Ball}\ and\ \citenamefont {James}(1987)}]{ball_1987}%
  \BibitemOpen
  \bibfield  {author} {\bibinfo {author} {\bibfnamefont {J.~M.}\ \bibnamefont
  {Ball}}\ and\ \bibinfo {author} {\bibfnamefont {R.~D.}\ \bibnamefont
  {James}},\ }\bibfield  {title} {\bibinfo {title} {Fine phase mixtures as
  minimizers of energy},\ }\href@noop {} {\bibfield  {journal} {\bibinfo
  {journal} {Archive for Rational Mechanics and Analysis}\ }\textbf {\bibinfo
  {volume} {100}},\ \bibinfo {pages} {13} (\bibinfo {year} {1987})}\BibitemShut
  {NoStop}%
\bibitem [{\citenamefont {Bhattacharya}(2003)}]{bhattacharya_2003}%
  \BibitemOpen
  \bibfield  {author} {\bibinfo {author} {\bibfnamefont {K.}~\bibnamefont
  {Bhattacharya}},\ }\href@noop {} {\emph {\bibinfo {title} {Microstructure of
  martensite: why it forms and how it gives rise to the shape-memory
  effect}}},\ Vol.~\bibinfo {volume} {2}\ (\bibinfo  {publisher} {Oxford
  University Press},\ \bibinfo {year} {2003})\BibitemShut {NoStop}%
\bibitem [{\citenamefont {Karami}\ \emph {et~al.}(2020)\citenamefont {Karami},
  \citenamefont {Zhu}, \citenamefont {Zeng}, \citenamefont {Tamura},
  \citenamefont {Yang},\ and\ \citenamefont {Chen}}]{karami_2020two}%
  \BibitemOpen
  \bibfield  {author} {\bibinfo {author} {\bibfnamefont {M.}~\bibnamefont
  {Karami}}, \bibinfo {author} {\bibfnamefont {Z.}~\bibnamefont {Zhu}},
  \bibinfo {author} {\bibfnamefont {Z.}~\bibnamefont {Zeng}}, \bibinfo {author}
  {\bibfnamefont {N.}~\bibnamefont {Tamura}}, \bibinfo {author} {\bibfnamefont
  {Y.}~\bibnamefont {Yang}},\ and\ \bibinfo {author} {\bibfnamefont
  {X.}~\bibnamefont {Chen}},\ }\bibfield  {title} {\bibinfo {title} {Two-tier
  compatibility of superelastic bicrystal micropillar at grain boundary},\
  }\href@noop {} {\bibfield  {journal} {\bibinfo  {journal} {Nano Letters}\
  }\textbf {\bibinfo {volume} {20}},\ \bibinfo {pages} {8332} (\bibinfo {year}
  {2020})}\BibitemShut {NoStop}%
\bibitem [{\citenamefont {Karami}\ \emph {et~al.}(2022)\citenamefont {Karami},
  \citenamefont {Chu}, \citenamefont {Zhu}, \citenamefont {Wang}, \citenamefont
  {Sun}, \citenamefont {Huang},\ and\ \citenamefont {Chen}}]{karami_2022}%
  \BibitemOpen
  \bibfield  {author} {\bibinfo {author} {\bibfnamefont {M.}~\bibnamefont
  {Karami}}, \bibinfo {author} {\bibfnamefont {K.}~\bibnamefont {Chu}},
  \bibinfo {author} {\bibfnamefont {Z.}~\bibnamefont {Zhu}}, \bibinfo {author}
  {\bibfnamefont {Z.}~\bibnamefont {Wang}}, \bibinfo {author} {\bibfnamefont
  {Q.}~\bibnamefont {Sun}}, \bibinfo {author} {\bibfnamefont {M.}~\bibnamefont
  {Huang}},\ and\ \bibinfo {author} {\bibfnamefont {X.}~\bibnamefont {Chen}},\
  }\bibfield  {title} {\bibinfo {title} {Orientation-dependent superelasticity
  and fatigue of {C}u{A}l{M}n alloy under in situ micromechanical tensile
  characterization},\ }\href@noop {} {\bibfield  {journal} {\bibinfo  {journal}
  {Journal of the Mechanics and Physics of Solids}\ }\textbf {\bibinfo {volume}
  {160}},\ \bibinfo {pages} {104787} (\bibinfo {year} {2022})}\BibitemShut
  {NoStop}%
\bibitem [{SMr()}]{SMref}%
  \BibitemOpen
  \href@noop {} {}\bibinfo {note} {See Supplemental Material at [URL will be
  inserted by publisher] for Methods, Figures S1 - S5, Legends of Movies S1 -
  S5.}\BibitemShut {Stop}%
\end{thebibliography}%

\end{document}